\documentclass{jps-cp}

\usepackage{txfonts}   
\usepackage{booktabs}
\usepackage{multirow}
\usepackage{float}
\usepackage{array}
\usepackage{multirow}

\newenvironment{acknowledgment}{\section*{Acknowledgment}}{}

\newcommand{\BETS}{$\lambda$-(BETS)$_2$GaCl$_4$}
\newcommand{\STF}{$\lambda$-(STF)$_2$GaCl$_4$}
\newcommand{\STFFe}{$\lambda$-(STF)$_2$FeCl$_4$}
\newcommand{\Wf}{$W_{\mathrm{frontier}}$}
\newcommand{\Ddim}{$\Delta_{\mathrm{dim}}$}
\newcommand{\EF}{$E_{\mathrm{F}}$}
\newcommand{\NEF}{$N(E_{\mathrm{F}})$}
\newcommand{\cnum}[1]{\textcircled{\scriptsize #1}}

\title{Wannier-Based Low-Energy Electronic Structure Analysis of \STF}

\author{Arifa D. \textsc{Agustin}$^{1,2*}$,
Aulia A. \textsc{Firdaus}$^{1}$, Fathan M. \textsc{Santana}$^{1}$,
Shuhei \textsc{Fukuoka}$^{3}$, Atsushi~\textsc{Kawamoto}$^{3}$,
Muhammad R. \textsc{Ramadhan}$^{4}$, Isao \textsc{Watanabe}$^{2,3}$,
Dita~P.~\textsc{Sari}$^{1,2,5}$,
and~Yasuyuki~\textsc{Ishii}$^{1}$}

\inst{$^{1}$Graduate School of Science and Engineering, Shibaura Institute of Technology (SIT), Saitama 337-8570, Japan\\
$^{2}$Nuclear Structure Research Group, RIKEN Nishina Center, Wako, Saitama 351-0198, Japan\\
$^{3}$Graduate School of Science, Hokkaido University, Sapporo 060-0810, Japan\\
$^{4}$Department of Chemical Engineering, UPN Veteran Yogyakarta, Yogyakarta 55281, Indonesia\\
$^{5}$Innovative Global Program, College of Engineering, SIT, Saitama 337-8570, Japan}

\email{na25101@shibaura-it.ac.jp}

\recdate{September 24, 2026}

\abst{The low-energy electronic structures of \BETS{} and \STF{} were investigated using density functional theory (DFT) and maximally localized Wannier functions (MLWFs). For \STF, DFT-selected and directly XRD-refined structural models were compared before and after fixed-lattice relaxation. Their initially different near-Fermi-level electronic structures become similar after relaxation. Relative to BETS, relaxed STF has a narrower frontier bandwidth while retaining a substantial dimerization energy scale. Two MLWFs were constructed for each selected unrelaxed structure to represent the upper frontier bands, with total spreads of 36.708~\AA$^2$ for BETS and 37.578~\AA$^2$ for STF. The Wannier-derived transfer integrals show a pronounced reduction of $t_{\mathrm{d1}}$ from 51~meV in BETS to 6.5~meV in STF, indicating selective modification of the low-energy electronic connectivity.}

\kword{Organic molecular conductors, \STF, density functional theory, maximally localized Wannier functions, dimerization}

\begin{document}
\maketitle

\section{Introduction}

Organic molecular conductors provide a platform for studying strongly correlated electron systems, where modest changes in molecular arrangement, composition, or pressure can alter the balance between kinetic energy and Coulomb interactions~\cite{Kobayashi1995,Tanaka1999,Kobayashi2004,Seo2004}. In $\lambda$-type conductors, donor-molecule dimerization defines key low-energy scales relevant to metallic, superconducting, and insulating states~\cite{Mori2001,Ito2022,Sawada2021,Fukuoka2024}. \BETS{} has been studied by extended H\"{u}ckel calculations~\cite{Montgomery1994} and DFT/Wannier approaches yielding molecular- and dimer-based low-energy models~\cite{Aizawa2018,Sari2021}. It is metallic at ambient pressure and becomes superconducting below about 5~K~\cite{Kobayashi1995,Tanaka1999}.

Replacing BETS with the asymmetric STF donor modifies the electronic structure while preserving the $\lambda$-type framework. Unlike \BETS, \STF{} remains insulating at ambient pressure, whereas metallicity and superconductivity emerge under pressure~\cite{Fukuoka2024,Minamidate2015}. This contrast motivates comparison of the relevant electronic energy scales~\cite{Minamidate2015,Agustin2026,Mori2002,Saito2018}, while the asymmetric S/Se distribution in STF raises the issue of structural-model sensitivity~\cite{Watanabe2020}.

In this work, we compare the electronic structures of \BETS{} and \STF{} using DFT. Two STF structural models, obtained by DFT configuration selection and direct XRD refinement, are examined before and after fixed-lattice relaxation. Their frontier bandwidths, dimerization gaps, and densities of states near the Fermi level are compared to assess robust BETS--STF trends and structural-model sensitivity. MLWFs for selected unrelaxed structures provide an initial low-energy description and comparison with the reported BETS network~\cite{Sari2021}.

\section{Methodology}

Single crystals of \BETS{} and \STF{} were synthesized by electrochemical oxidation. Single-crystal X-ray diffraction (XRD) measurements were performed at 300~K using a Bruker D8 QUEST diffractometer at the Techno Plaza, Shibaura Institute of Technology. Data processing and structural refinement were performed using APEX5, and the resulting crystal structures were visualized using VESTA.

Three structural models were used: the 300~K XRD-refined BETS structure and two STF models. The first STF model was derived from isostructural \STFFe{} by replacing Fe with Ga and selecting the lowest-energy S/Se configuration among those examined ($E_0 = -143.24357$~eV)~\cite{Watanabe2020}. The second model was obtained directly from XRD refinement. DFT calculations were performed using VASP~\cite{Kresse1996} with the PAW method~\cite{Kresse1999}, the PW91--GGA functional~\cite{Perdew1992}, a plane-wave energy cutoff of 500~eV, and a $4\times4\times4$ Monkhorst--Pack $k$-point mesh. The electronic convergence criterion was $10^{-7}$~eV, and the DOS used 10,000 energy-grid points. Experimental geometries and fixed-lattice relaxed structures (ISIF = 2) were calculated, with internal coordinates optimized to residual forces below 0.01~eV~\AA$^{-1}$. Calculations were performed using the RIKEN HOKUSAI and JHPCN Miyabi-C supercomputing systems.

The low-energy electronic structures were characterized by the frontier bandwidth \Wf{} (width of the two upper frontier bands), dimerization gap \Ddim{} (separation between the upper and lower frontier-band pairs), density of states \NEF, and $\eta = W_{\mathrm{frontier}}/\Delta_{\mathrm{dim}}$. MLWFs were constructed with Wannier90~\cite{Marzari1997,Mostofi2014} for the unrelaxed BETS XRD-refined and STF DFT-selected structures. Two Wannier functions per compound represent the upper frontier bands within the four-band frontier manifold.

\section{Results and Discussion}

\begin{figure}[tbh]
\centering
\includegraphics[width=1\linewidth]{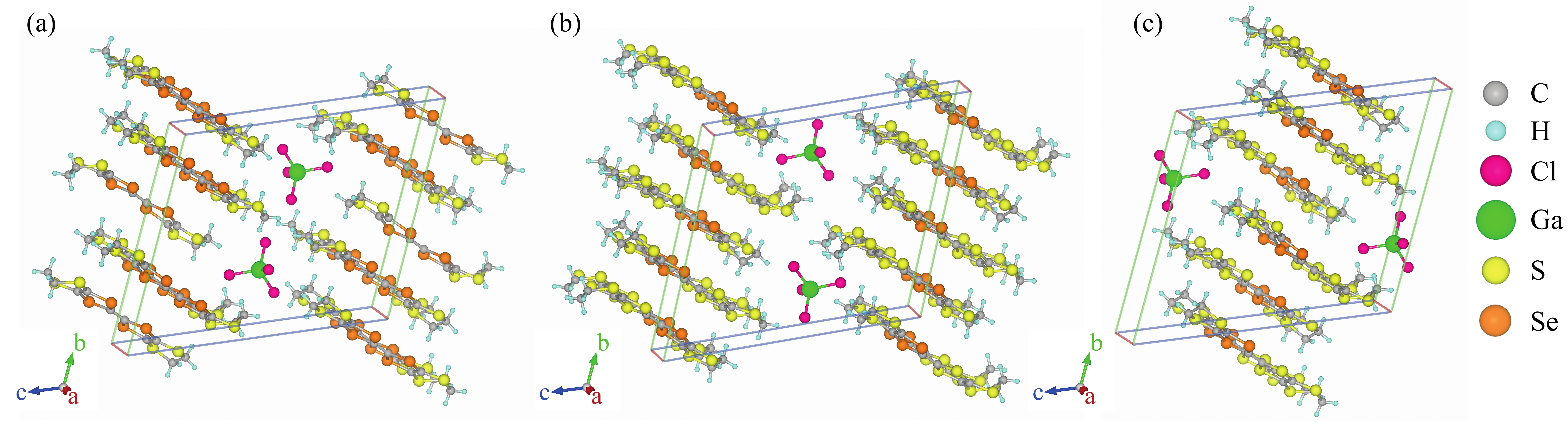}
\caption{Crystal structures of (a) \BETS{} obtained from XRD refinement, (b) \STF{} based on the DFT-selected S/Se configuration, and (c) \STF{} obtained from direct XRD refinement. The structure in (b) was derived from the isostructural \STFFe{} structure by replacing Fe with Ga, followed by selection of the lowest-energy S/Se configuration among those examined by DFT calculations.}
\label{fig:1}
\end{figure}

Figure~\ref{fig:1} compares the three structural models, with crystallographic parameters listed in Table~\ref{tab:1}. The XRD-refined BETS structure has a unit-cell volume of 1786.04(9)~\AA$^3$, while the DFT-selected and XRD-refined STF models have slightly smaller volumes of 1746.75 and 1758.0(6)~\AA$^3$, respectively. Despite their different origins, the two STF models retain similar lattice parameters and the characteristic $\lambda$-type molecular arrangement.

\begin{table}[H]
\caption{Structural parameters of the initial 300~K structural models used in the DFT calculations.}
\label{tab:1}
\centering
\fontsize{10pt}{10pt}\selectfont
\setlength{\tabcolsep}{17pt} 
\renewcommand{\arraystretch}{1}

\begin{tabular}{cccc}
\toprule
Parameter & \BETS & \multicolumn{2}{c}{\STF} \\
\cmidrule(lr){1-1}\cmidrule(lr){2-2}\cmidrule(lr){3-4}
Structure origin & XRD-refined & DFT-selected configuration & XRD-refined \\
\midrule
$a$ (\AA) & 6.6096(2) & 6.5645 & 6.5805(13) \\
$b$ (\AA) & 16.1821(4) & 16.2020 & 16.219(3) \\
$c$ (\AA) & 18.6205(5) & 18.286 & 18.361(4) \\
$\alpha$ ($^\circ$) & 112.584(10) & 112.347 & 112.422(7) \\
$\beta$ ($^\circ$) & 98.356(10) & 98.443 & 98.371(7) \\
$\gamma$ ($^\circ$) & 96.789(10) & 96.799 & 97.027(7) \\
$V$ (\AA$^3$) & 1786.04(9) & 1746.75 & 1758.0(6) \\
\bottomrule
\end{tabular}

\vspace{3pt}
\begin{minipage}{\linewidth}
\footnotesize Values in parentheses are estimated standard uncertainties from the XRD refinement. Lattice parameters of the DFT-selected model were taken from \STFFe~\cite{Watanabe2020} and are given without uncertainties.
\end{minipage}
\end{table}

\begin{figure}[H]
\centering
\includegraphics[width=0.85\linewidth]{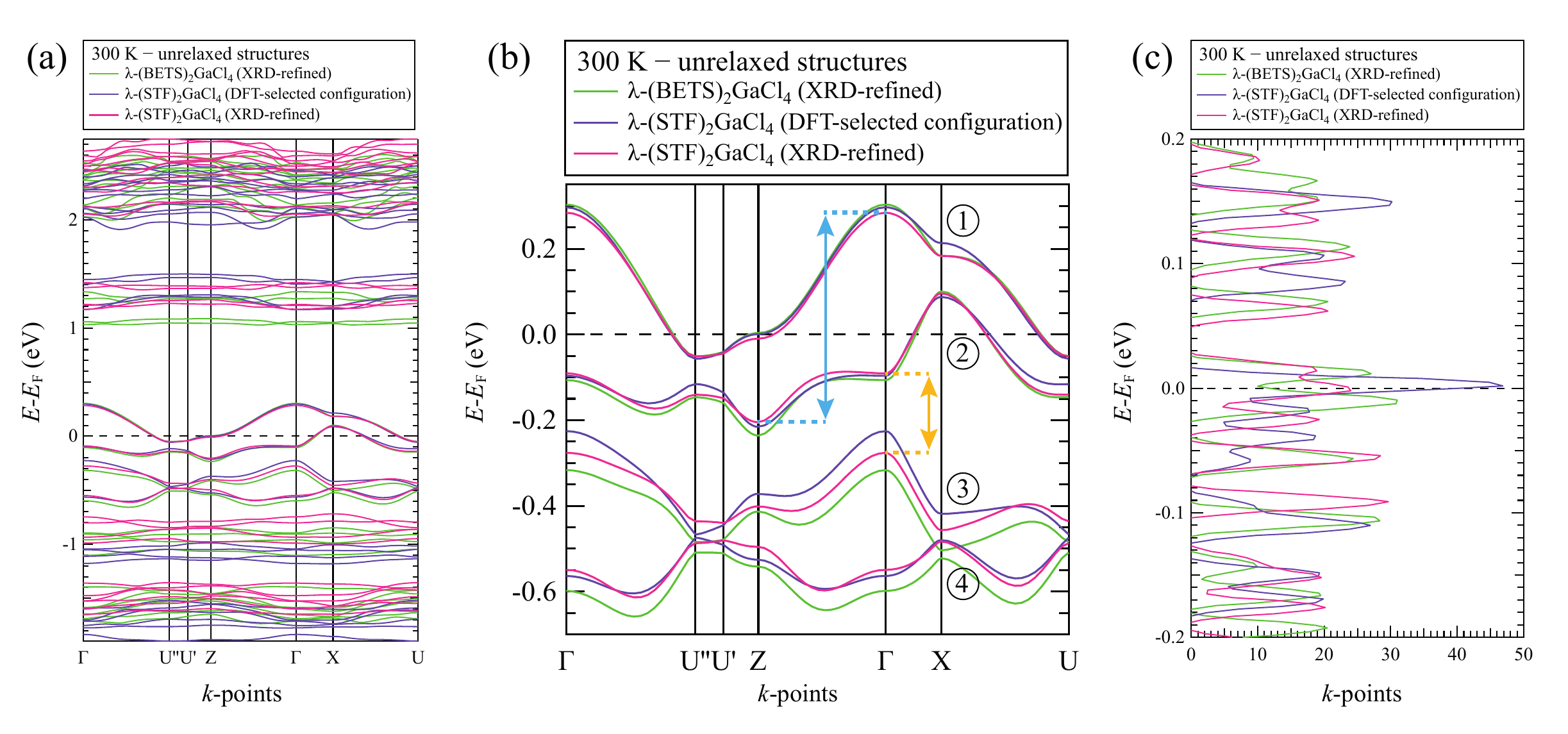}
\caption{DFT electronic structures of \BETS{} and \STF{} calculated using the unrelaxed 300~K structural models: (a) full band structures, (b) enlarged frontier-band region, and (c) DOS near the Fermi level. Blue and yellow arrows in (b) indicate \Wf{} and \Ddim, respectively, while \cnum{1}--\cnum{4} label the frontier bands.}
\label{fig:2}
\end{figure}

\begin{figure}[H]
\centering
\includegraphics[width=0.85\linewidth]{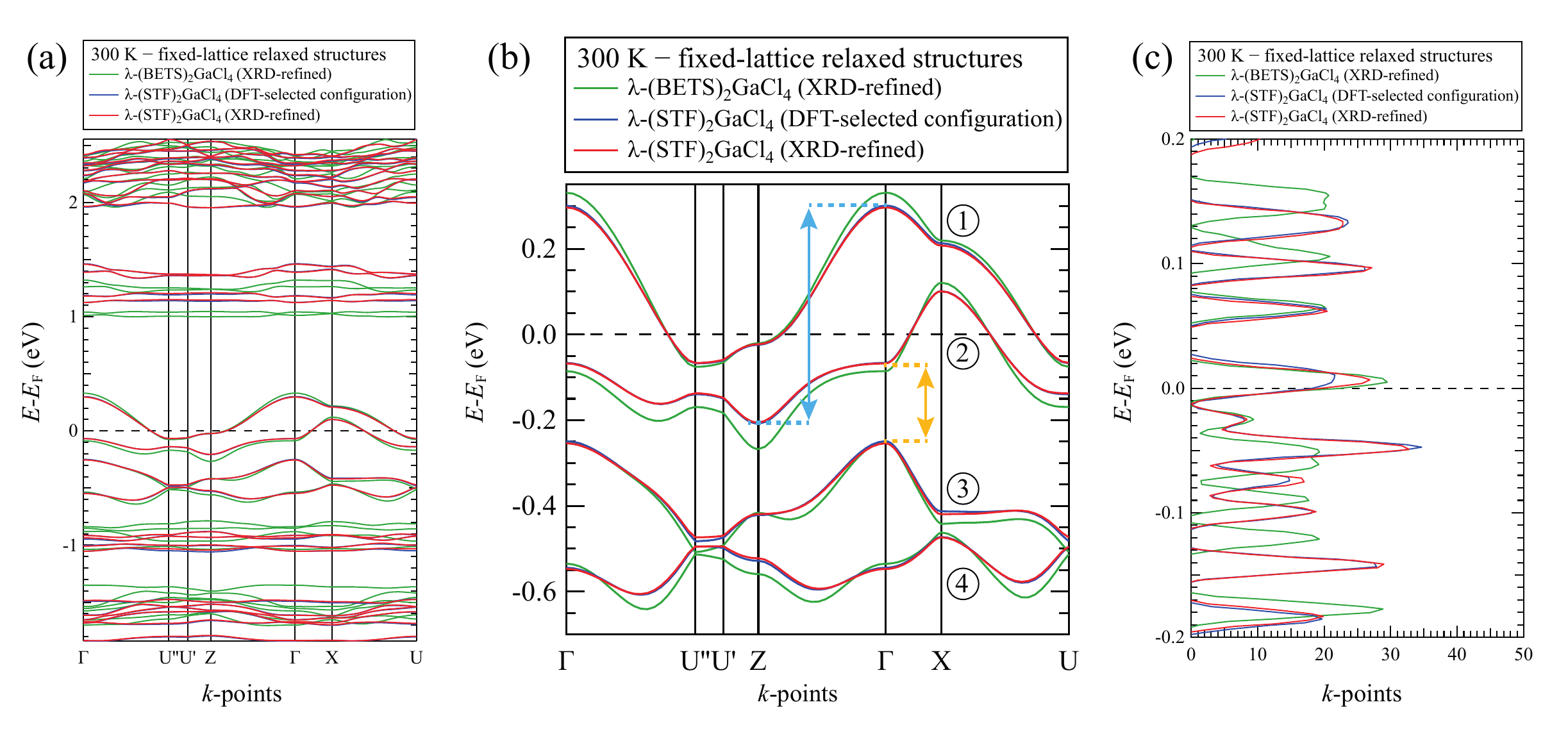}
\caption{DFT electronic structures of \BETS{} and \STF{} after fixed-lattice structural relaxation of the 300~K structural models: (a) full band structures, (b) enlarged frontier-band region, and (c) DOS near the Fermi level. Arrows and band labels are as in Fig.~\ref{fig:2}(b).}
\label{fig:3}
\end{figure}

\vspace{10pt}

Figure~\ref{fig:2} compares the unrelaxed 300~K electronic structures. All three models show a similar four-band frontier manifold, with the two upper bands crossing \EF{} in the present GGA description. For \STF, this metallic Kohn--Sham band structure should not be identified with the experimentally observed low-temperature insulating state~\cite{Fukuoka2024,Minamidate2015}, because the relevant correlation effects are not included. The bands are therefore used to compare the underlying single-particle bandwidth and dimerization scales.

As summarized in Table~\ref{tab:params}, both unrelaxed STF models have narrower frontier manifolds than BETS, consistent with a reduced kinetic-energy scale. This trend is qualitatively consistent with STF lying closer to the insulating side of the bandwidth-controlled phase diagram~\cite{Mori2001,Ito2022}, although quantitative comparison of correlation strength also requires an interaction scale. The two STF models nevertheless differ in \Ddim, $\eta$, and particularly \NEF, showing sensitivity to the initial structural model.

\begin{table}[H]
\caption{Characteristic electronic-structure parameters of \BETS{} and \STF.}
\label{tab:params}
\centering

\fontsize{9pt}{9pt}\selectfont
\setlength{\tabcolsep}{13pt}
\renewcommand{\arraystretch}{1.3}

\begin{tabular}{
>{\raggedright\arraybackslash}m{1.5cm}
>{\raggedright\arraybackslash}m{3.2cm}
>{\centering\arraybackslash}m{1.0cm}
>{\centering\arraybackslash}m{1.0cm}
>{\centering\arraybackslash}m{1.5cm}
>{\centering\arraybackslash}m{1.5cm}
}
\toprule
\shortstack[l]{Structural\\state}
& Structure
& \shortstack{\Wf\\(eV)}
& \shortstack{\Ddim\\(eV)}
& \shortstack{$\eta =$\\$W_{\mathrm{frontier}}/\Delta_{\mathrm{dim}}$}
& \shortstack{\NEF\\(states/eV/cell)} \\
\midrule

\multirow{3}{*}{Unrelaxed}
& BETS, XRD-refined & 0.539 & 0.210 & 2.565 & 11.69 \\
& \shortstack[l]{STF, DFT-selected\\configuration}
& 0.512 & 0.129 & 3.967 & 40.63 \\
& STF, XRD-refined & 0.488 & 0.185 & 2.636 & 23.72 \\

\midrule

\multirow{3}{*}{\shortstack[l]{Fixed-lattice\\relaxed}}
& BETS, XRD-refined & 0.598 & 0.164 & 3.639 & 22.66 \\
& \shortstack[l]{STF, DFT-selected\\configuration}
& 0.508 & 0.183 & 2.783 & 18.30 \\
& STF, XRD-refined & 0.502 & 0.186 & 2.693 & 19.72 \\

\bottomrule
\end{tabular}
\end{table}

\begin{figure}[H]
\centering
\includegraphics[width=1\linewidth]{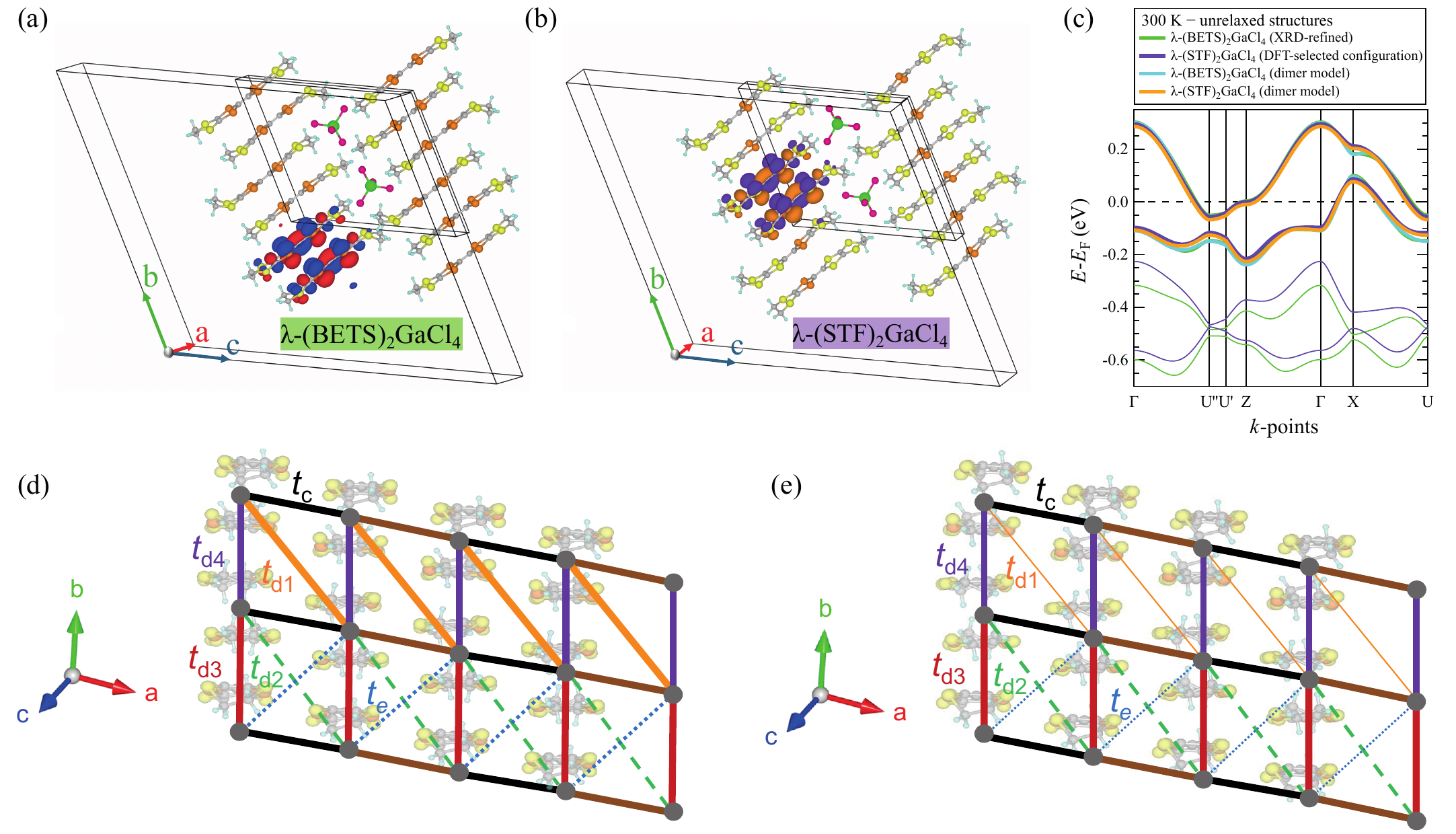}
\caption{(a,b) MLWFs for the two upper frontier bands of unrelaxed XRD-refined \BETS{} ($\Omega_{\mathrm{BETS}} = 36.708$~\AA$^2$) and DFT-selected configuration \STF{} ($\Omega_{\mathrm{STF}} = 37.578$~\AA$^2$), respectively. (c) DFT and Wannier-interpolated bands for both compounds. (d,e) Dimer-based transfer-integral network for BETS and STF, respectively. Gray circles indicate the centers of donor-molecule dimers, and the colored lines represent $t_{\mathrm{c}}$, $t_{\mathrm{d1}}$, $t_{\mathrm{d2}}$, $t_{\mathrm{d3}}$, $t_{\mathrm{d4}}$, and $t_{\mathrm{e}}$ as listed in Table~\ref{tab:transfer}.}
\label{fig:wannier}
\end{figure}

Fixed-lattice relaxation reduces these differences (Fig.~\ref{fig:3}). The two STF models yield \Wf{} = 0.508 and 0.502~eV and \Ddim{} = 0.183 and 0.186~eV, compared with 0.598 and 0.164~eV for BETS. Thus, STF has a narrower frontier bandwidth while retaining a substantial dimerization energy scale. The relaxed structures remain consistent with a dimerized low-energy description. These single-particle results do not determine the many-body magnetic ground state.

As a first step toward constructing a reduced low-energy representation, two MLWFs were constructed for each unrelaxed structure: the BETS XRD-refined structure and the STF DFT-selected configuration. Figures~\ref{fig:wannier}(a) and \ref{fig:wannier}(b) show representative orbital distributions with appreciable weight on donor-molecule pairs. The total spreads are 36.708~\AA$^2$ for BETS and 37.578~\AA$^2$ for STF. The Wannier-interpolated bands closely reproduce the two upper DFT frontier bands [Fig.~\ref{fig:wannier}(c)], supporting the transfer-integral analysis. Using the labeling convention of Ref.~\cite{Sari2021}, Figs.~\ref{fig:wannier}(d) and \ref{fig:wannier}(e) show six interdimer pathways, $t_{\mathrm{c}}$, $t_{\mathrm{d1}}$, $t_{\mathrm{d2}}$, $t_{\mathrm{d3}}$, $t_{\mathrm{d4}}$, and $t_{\mathrm{e}}$. The calculated BETS magnitudes broadly agree with Ref.~\cite{Sari2021}. The signs of the transfer integrals depend on the phase convention of the MLWFs. In STF, $t_{\mathrm{d1}}$ decreases markedly from 51 to 6.5~meV, while $t_{\mathrm{c}}$, $t_{\mathrm{d3}}$, and $t_{\mathrm{d4}}$ remain comparable. This selective modification of transfer pathways indicates that replacing BETS with STF changes the electronic connectivity rather than uniformly reducing all hopping amplitudes. This change may reflect modified local interdimer geometry and orbital overlap associated with the S/Se arrangement, although its microscopic origin remains open.

\begin{table}
\caption{Calculated transfer integrals (meV) for the unrelaxed \BETS{} XRD-refined and \STF{} DFT-selected structures, compared with the previously reported BETS values~\cite{Sari2021}.}
\label{tab:transfer}
\centering
\fontsize{10pt}{10pt}\selectfont
\setlength{\tabcolsep}{19pt}
\renewcommand{\arraystretch}{1}

\begin{tabular}{
>{\centering\arraybackslash}m{1.5cm}
>{\centering\arraybackslash}m{2.5cm}
>{\centering\arraybackslash}m{2.5cm}
>{\centering\arraybackslash}m{3.5cm}
}
\toprule
Label
& \shortstack[c]{\BETS{}\\$[11]$} 
& \shortstack[c]{\BETS{}\\(this work)}
& \shortstack[c]{\STF{}\\DFT-selected configuration} \\
\midrule

$t_{\mathrm{c}}$  & 64 & 58 & 56 \\
$t_{\mathrm{d1}}$ & 52 & 51 & 6.5 \\
$t_{\mathrm{d2}}$ & 13 & $-14$ & 13 \\
$t_{\mathrm{d3}}$ & 76 & 75 & 75 \\
$t_{\mathrm{d4}}$ & 64 & 68 & 60 \\
$t_{\mathrm{e}}$  & $-17$ & 17 & $-9.8$ \\

\bottomrule
\end{tabular}
\end{table}

\section{Conclusion}

DFT and dimer-based Wannier analyses were used to compare \BETS{} and \STF. Structural relaxation reduces the electronic-structure differences between the two STF models. Relative to BETS, relaxed STF has a narrower frontier bandwidth while retaining a substantial dimerization energy scale. Wannier analysis of the unrelaxed structures further shows a pronounced reduction of $t_{\mathrm{d1}}$ from 51~meV in BETS to 6.5~meV in STF, while several other interdimer couplings remain comparable. This modification may be related to the S/Se arrangement, although its microscopic origin remains open. The results indicate a selective change in the low-energy electronic connectivity rather than a uniform reduction of the interdimer transfer integrals.

\begin{acknowledgment}
We acknowledge T. Kobayashi of Hokkaido Univ., T. Koretsune of Tohoku Univ., Y. Nagai of the Univ. of Tokyo, and H. Seo of RIKEN for helpful discussions. ADA is supported by the SIT scholarship for foreign graduate students. DPS and IW are supported by JSPS KAKENHI Grant Nos. JP21K13885, JP24K06133, JP23H04529, and JP25H01559, the JSPS Asia-Africa Core-to-Core Program No. JPJSCCB20240005, and JST NEXUS Y-tec Nos. Y2024L0906032 and Y2025L1015059. SF is supported by JSPS KAKENHI Grant No. JP22K03503, the Izumi Science and Technology Foundation, and the JGC-S Scholarship Foundation. Computations were performed on the RIKEN HOKUSAI (IDs: RB230087, RB260021) and JHPCN Miyabi-C (ID: jh260055) supercomputers.
\end{acknowledgment}

\end{document}